# Activity Detection from Wearable Electromyogram Sensors using Hidden Markov Model

Rinki Gupta, Karush Suri*
Electronics and Communication Engineering Department
Amity University Noida, Uttar Pradesh-201313, India
rgupta3@amity.edu, karushsuri@gmail.com

*Abstract*—Surface electromyography (sEMG) has gained significant importance during recent advancements in consumer electronics for healthcare systems, gesture analysis and recognition and sign language communication. For such a system, it is imperative to determine the regions of activity in a continuously recorded sEMG signal. The proposed work provides a novel activity detection approach based on Hidden Markov Models (HMM) using sEMG signals recorded when various hand gestures are performed. Detection procedure is designed based on a probabilistic outlook by making use of mathematical models. The requirement of a threshold for activity detection is obviated making it subject and activity independent. Correctness of the predicted outputs is asserted by classifying the signal segments around the detected transition regions as activity or rest. Classified outputs are compared with the transition regions in a stimulus given to the subject to perform the activity. The activity onsets are detected with an average of 96.25% accuracy whereas the activity termination regions with an average of 87.5% accuracy with the considered set of six activities and four subjects.

*Keywords— sEMG; activity detection; posterior states; Viterbi algorithm; HMM; stimulus*

## I. Introduction

Study of various hand gestures has gained significant importance in the modern age [1, 2]. Amputee treatment, digital gesture recognition, healthcare development and sign language communication are some of the few applications of this diverse field. Hand gestures are analyzed as surface electromyogram (sEMG) signals which are obtained from the electrical action of muscle tendons in the limb occurring due to motor functionality. Generated signals are recorded with the help of wearable sensors consisting of surface compatible electrodes capable of sensing electrical activity. The need for complex medical equipment and chemical requirement for diagnostic treatment is overcome with the provision of sEMG electronics.

A gesture recognition system requires detailed analysis of the gesture to be performed. These require various activity detection algorithms which predict the state of the motor muscles [2-4]. Predictions are based on hand kinematics, muscle sensing, supervised algorithms used and various other factors. Muscle activity regions may simply be detected by monitoring when the raw sEMG signal or an average of sEMG signals obtained from different sEMG sensors placed on the hand exceed a predefined threshold [2]. Although this approach is computationally less expensive, however, the accuracy of detection depends on the selection of a suitable threshold for the subject and activity under consideration. Similarly, using other parameters, such as root mean square (RMS) value of the signal or high-order statistical parameters also requires the selection of a threshold for activity detection [3]. In [4] an approach for offline detection of hand activities where the threshold required for activity detection is not as sensitive to amplitude variation in sEMG signals as in the case of RMS of the signal.

In this paper, an activity detection approach is proposed for detecting the periods of activity and rest from sEMG signals using Hidden Markov Models (HMM). HMM has often been employed for detection of voice activity [5]. The proposed approach processes the raw signals in time domain using HMM to detect the possible regions of hand activity, which are further refined using logical constraints to determine the activity onset and termination instants. The algorithm does not require the use of a threshold, which makes it versatile to detect various activities from sEMG signals recorded for different subjects.

Organization of the paper is as follows. Section II contains a brief review of the techniques used in this work. Section III describes the experimental setup used to record the dataset considered in this work. Section III also contains the detailed description of the proposed hand activity detection approach and the classification technique used for validating the results to assess the performance of the same. Results depicting the suitability of proposed algorithm for activity detection and the classification of regions around the detected transitions are given in Sec. IV. Section V concludes the paper.

## II. Review of HMM and its Applications

In order to study the hand gestures using sEMG, one must have a good idea regarding when the limb is in motion. Performing a gesture consists of two phases: 1) the activity phase and 2) the rest phase. A proper distinction between the two phases is essential for gesture analysis. HMMs are useful in activity detection because they exploit the temporal dependence of system states under a probabilistic framework. HMMs are similar to Markov chain models with the only difference being that in an HMM, the states are unknown or *hidden* [5, 6]. These states are predicted on the basis of initial state, transition probabilities and observations obtained from the signal. For activity detection in speech, video or sEMG signals, the states are either rest or activity. Also, in case of activity detection, it is assumed that the initial state ($\pi_i$) is most likely to be the rest state since that is the natural order in which a human would perform the activity. Transition probabilities is the probability of a state $q_i$ transitioning to another state $q_j$, where $i$ and $j$ are the instances of the hidden states at a given time instant, $t$. If the transition probability depends on only the previous state in time, its referred to as a 1st order HMM. Once a state is predicted at any given instant $t$, an observation symbol $o_k$ is associated to it, where $k$ denotes the total number of symbols from 1 to $M$. This gives rise to another set of probabilities known as emission probabilities which is defined as the probability that the output is the symbol $o_k$ given the current state is $q_i$. Mathematically, the set of probabilities are expressed as follows-

$$\text{Hidden States, } Q = \{q_i\}, \quad i = 1,2,\dots N \quad (1)$$

$$\text{Transition Probabilities, } T_{ij} = P(q_j \text{ at } t+1 \mid q_i \text{ at } t) \quad (2)$$

$$\text{Symbols, } O = \{o_k\}, \quad k = 1,2,\dots M \quad (3)$$

$$\text{Emission Probabilities, } E_{ik} = P(o_k \mid q_i)\} \quad (4)$$

Depending on the transitions possible between states in an HMM, the topology of HMM may be classified as left-to-right and ergodic HMM [6,7]. Here, a 1st-order left-to-right HMM, as depicted in Fig. 1, has been considered because in activity detection, the transitions progress in time.

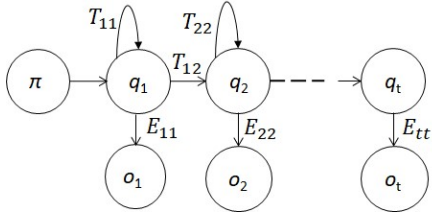

Fig. 1. HMM topology for activity detection

Computation of the state sequence is carried out by making use of the Viterbi algorithm [6,7]. The algorithm chooses the best state sequence which optimizes the likelihood of the state sequence corresponding to the given observation sequence. Let $\Theta_t(i)$ be the maximum probability of state sequences having length $t$ that end with state $i$ and generate the $t$ first observations for the model. This is expressed mathematically in (5).

$$\Theta_t(i) = max\{P(q(1), q(2), \dots q(t-1); o(1), o(2), \dots o(t) \mid q(t) = q_i)\} \quad (5)$$

HMM has been employed in voice activity detection in case of speech signals [5] as well as for motion detection in videos [8]. These models possess a fast-computational procedure and are state based. Prediction of the output states is based on guessing the hidden states for the input sequence. HMM has also been used with sEMG signals recorded from limbs or other parts of the body. In [9], signals recorded from accelerometers worn in a waist-belt by the subjects are processed using HMM to identify usual daily activities. In [10], speech recognition system is designed using sEMG signals recorded by sensors mounted of subjects' fingers, which are pressed against the face of the subject while he speaks. Here, HMM is applied on features extracted from sEMG signals to identify the speech sounds. These features improve the performance of models by reducing correlation between various instances of the signal. However, feature extraction and selection for state based prediction is inconvenient due to complex computations. Linking the similarity between voice and sEMG signals on dimensional lines, HMMs provide a convenient algorithm for activity detection.

HMM has been used identify regions as neither belonging to any specified activity or rest, so as to improve the classification accuracy provided by SVM by removing the falsely detected states at transition regions [7]. Another approach would be to detect the activity in the signal and then validate the detected instances by treating it as a binary classification problem. This approach has been used in this paper and is described as follows. First, the sEMG signals are labeled on the basis of a pre-defined hypothesis or the initial states indicating the periods of activity and rest. Then, the features extracted from the input signals are used to predict the statistically most probable states using the HMM approach. Finally, the features corresponding to rest and activity phases as indicated by the predicted states are compared with those indicated by the original sequence by using them to classify the activity phase from the rest phase; hence, validating the performance of the activity detection algorithm. Some of the commonly used classifiers in are k-Nearest Neighbor (kNN), Random Forest (RF) and Support Vector Machines (SVM) [6]. An important advantage of SVM over other algorithms is the computational efficiency provided as a result of the kernels adopted in the high dimensional feature space [11]. Hence, SVM has been used in this paper to validate the results obtained from the proposed algorithm.

III. PROPOSED HAND ACTIVITY DETECTION USING HMM

A. The sEMG dataset

The sEMG signals considered in this work have been recorded from the surface of skin over three muscles namely, Flexor Capri Ulnaris and Extensor Capri Radialis and Brachioradialis on the right arm of the subject. The objective is to utilize the sEMG signals to detect the regions of activity and rest in the 6 gestures performed by the right hand. The sEMG data has been collected from 4 healthy subjects in the age group of 22-30 years, all right-handed females. The six gestures that are performed are signs for Win, Loose, Key, Bold, Confident and Sorry from the Indian Sign Language published by the Faculty of Disability Management and Special Education (FDMSE) of Ramakrishna Mission Vivekananda University (RKMVU), Coimbatore, India. A stimulus of 3 seconds is played to the subject to indicate when the gesture is to be performed, with 5 seconds of rest in between each gesture. Each subject performed 20 repetitions for each gesture, the sEMG signals of which are recorded in continuation as one recording. A 2 min rest is given in between two recordings to avoid muscle fatigue. The sEMG signals are recorded using the Delsys wireless EMG system. The sEMG signals are obtained with a sample rate of 1.1 kHz and a bit depth of 16 bits. A delay of 0.5 seconds has also been taken into account generated by the recording apparatus and compensated as hardware delay. Fig. 2 shows the complete experimental setup used in the recording of sEMG signals.

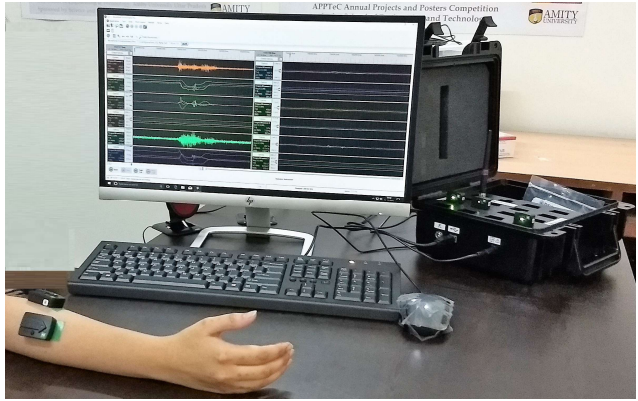

Fig. 2. Experimental setup for recording of sEMG signals

Due to delayed response of the subject, the activity onset instant generally does not coincide with the instant at which the stimulus starts to play. Moreover, the gesture may not require 3 seconds to perform and the gesture termination instant is generally less as compared to when the stimulus ends. Hence, accurate detection of onset and termination instants of the gesture from the sEMG signals is an important step before processing the signal for enhancement or classification.

*B. Proposed Hand Activity Detection using HMM*

The proposed hand activity detection process is depicted in Fig. 3. Applying the state based model on sEMG signals requires an initial guess of the sequence of states. The initial guess of the states is obtained from the stimulus played during the recording of sEMG signals. When the audio stimulus is played, the state is set to 1 to indicate possible activity region, else the state is set to 0 to indicate possible rest region. The RMS of sEMG signal is quantized to 16 levels to reduce the computational complexity of the detection algorithm. The quantized RMS signals are fed as inputs to HMM, repetition by repetition. Corresponding to each repetition, estimation of transition and emission probabilities ($T_{ij}$ and $E_{ik}$, respectively) takes place which are used to train the model. Once the model is trained, calculation of posterior states $\Theta_t(i)$ is carried out using the Viterbi algorithm, which results in the detected sequence.

The detected HMM states may contain certain spurious transition between rest and activity states, particularly at activity termination instant where the signal smoothly decays into noise recorded during the rest regions. Hence, certain logical constraints are used to decide the instants of activity onset and termination. Here, minimum distance constraint has been applied such that a detected activity region of duration less than 0.8 seconds is considered spurious, since none of the signs could be performed in less than a second. Also, if multiple onset instants are detected, the first instant of onset is considered as the valid onset instant, while for activity termination, the last instant is considered to be the valid instant. Finally, the detected sequence is compared to the original sequence for falsely detected points of rest and activity. The wrongly detected points are verified by means of SVM classification as discussed in the next sub-section.

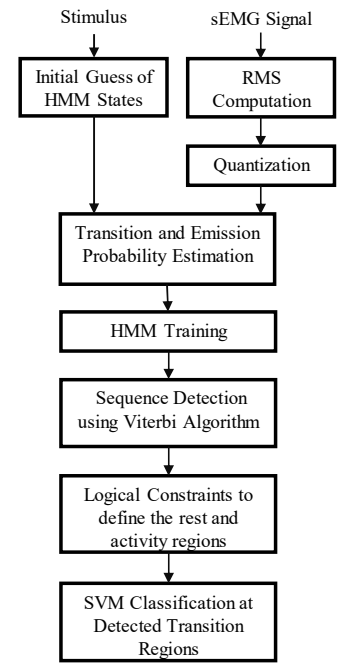

Fig. 3. Proposed hand activity detection algorithm

*C. Validation using SVM Classification*

In order to assess the correctness of detection, the conflicting instances are classified as activity or rest. These data points are accumulated at the transition regions, which is when the rest period transits into activity period and vice-versa. These two regions correspond to the edges at which the activity period begins and ends respectively and are labeled as starting edge and ending edge. For classifying the data points into two categories of rest and activity at the starting and ending edges, a 0.5 second window is generated for each repetition. The window is centered at the respective edge, considering 0.25 seconds of rest on the left side and 0.25 seconds on the right side. Scattering of these data points is studied which provides a concise indication regarding the grouping of rest and activity points. These points are accumulated in the form of clusters. However, points which are falsely detected are placed at certain distances from their respective clusters. In order to assess the true value of these points, classification is performed using SVM and compared with the original scattering of the digital stimulus.

The SVM makes use of the most optimized hyper-plane in the high- dimensional feature space, thus mapping the inputs implicitly to their corresponding outputs. The kernel trick is applied, wherein a kernel, which is a generalized form of the function is used. This allows the function to solve the entities in a dimensionally improved feature space by successive computation making use of the dot product. Due to its non-probabilistic functionality, a clear margin is drawn between the classes to which the groups belong and this provides a definite value for each point. In case of binary classification, the predicted values are consequence of the two classes. The two classes of rest and activity are marked with indicators '0' and '1' respectively. These are distinguished between on the basis of RMS computation from the three channels which act as features for the data points. SVM is firstly trained with half of the data points. The training is then tested on the validation set which consists of the remaining half of the data points. Thus, a well- partitioned data is provided for testing which is a significant step in eliminating correlation and avoiding over-fitting problems.

Classification approach of the sEMG data points provides a clear distinction between the instances of activity and rest in case of HMM outputs. Conflicting instances are grouped as activity or rest, depending upon their RMS values. The scattering of classified outputs is compared to the original scattering of the RMS values. A clear verification for the predicted values is provided by observing a similar grouping of the similar data points on the scatter plots.

## IV. RESULTS AND DISCUSSION

Figure 4a shows an instance of the sEMG signal, its RMS envelope and the stimulus, which has been scaled for viewability. The detected HMM states, as shown in Fig. 4b, have rapid transition between rest and activity states, at the transition regions. For the HMM states shown in Fig. 4b, the resulting activity and rest regions after applying the logical constraints are shown in Fig. 4c. As compared to the stimulus in Fig. 4a, the detected activity and rest regions in Fig. 4c are more accurate.

In order to observe the data points at starting and ending edges, scatter plots are studied for RMS values extracted from segments around the edges of the original stimulus and the predicted activity regions, as explained in Section IIIc. The results for the sign 'Sorry' for all the four subjects are shown in Fig. 5. In the case of original values (Figs. 5a and 5c), the clustering is not prominent as clusters tend to overlap each other due to the absence of a clear distinction. Thus, the stimulus- based assumption does not provide a clear distinction

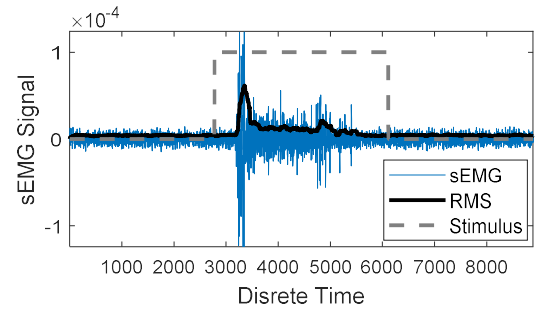

(a) sEMG signal, RMS envelope and Stimulus

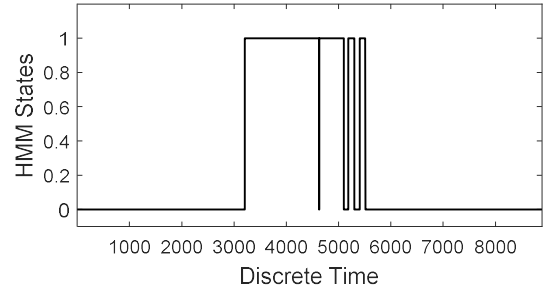

(b) HMM States

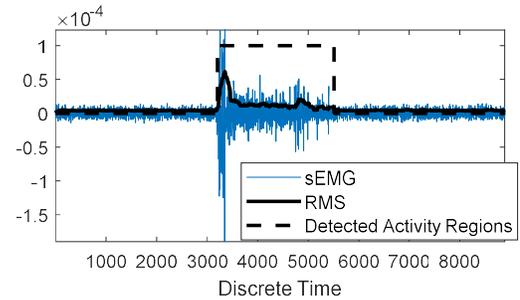

(c) sEMG signal, RMS envelope and the detected activity region from HMM states

Fig. 4. Detected Activity and Rest regions for a repition of the sign for Sorry

between the regions of activity and rest. On the other hand, a clear clustering effect is observed in the scattering of detected outputs (Figs. 5b and 5d). The two clusters of activity and rest are at a certain distance in the RMS feature space. The clustering can be further assessed by making use of a classification algorithm which would provide a clear demarcation between rest and activity data points.

For a proper validation of the detected results, the data points at the starting and ending edges are classified using SVM. As shown in Fig. 6, the classification accuracies of RMS values from the segments around the detected activity onset

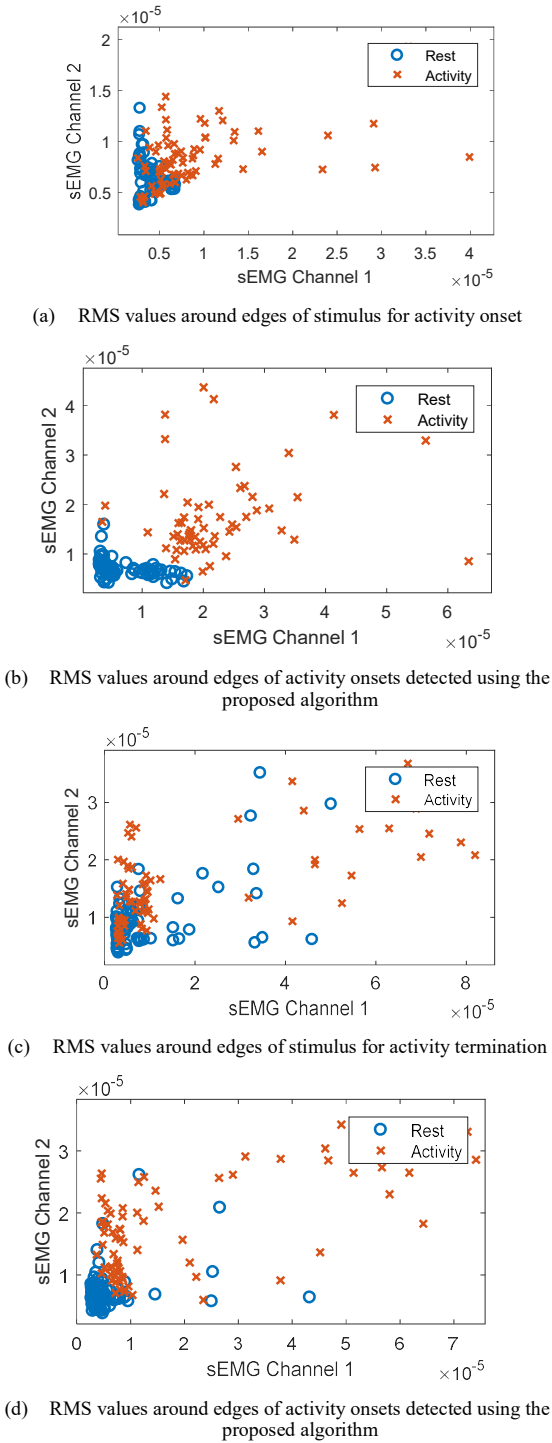

(a) RMS values around edges of stimulus for activity onset

(b) RMS values around edges of activity onsets detected using the proposed algorithm

(c) RMS values around edges of stimulus for activity termination

(d) RMS values around edges of activity onsets detected using the proposed algorithm

Fig. 5. Scatter plot of RMS values for all subjects for the sign 'Sorry'

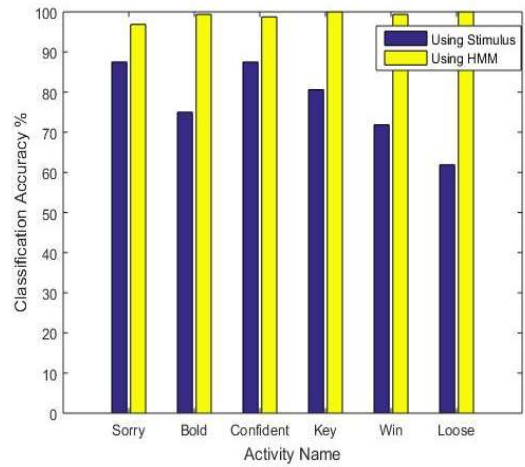

(e) For activity onset

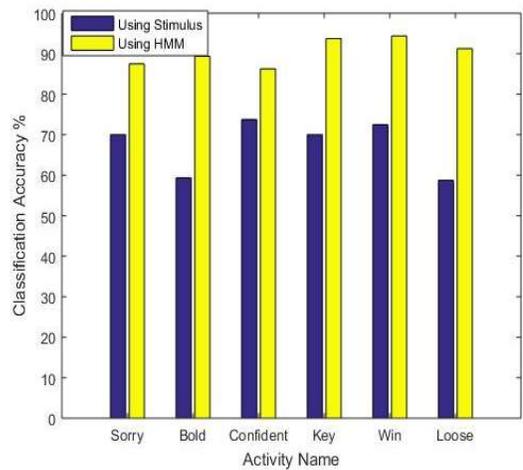

(f) For activity termination

Fig. 6. Comparison of clasification accuracy values for original RMS instances vs. HMM predicted outputs

and termination regions using the proposed HMM based algorithm (shown in light colour) are compared to that for the original RMS values (shown in dark colour). Original values for each gesture in case of each subject are classified on the basis of RMS features from the three channels. The comparison drawn is assessed on the basis of classification accuracy which is a significant measure to compute the percentage of true positive outputs obtained from the input data. As depicted in Fig 6(a) and Fig 6(b), for each gesture, the classification accuracies are greater in the case of HMM outputs when compared to the original RMS values for both starting and ending edges. For each gesture, the increased amount of correlation is removed from the HMM output data points thus, providing an improved procedure for activity detection. The average classification accuracy for activity onset and termination instants are 96.25% and 88.13%, respectively for the proposed algorithm as compared to 87.50% and 71.25%, respectively for the original stimulus for 5-fold cross validation, hence validating the effectiveness of the proposed algorithm for activity detection.

## V. CONCLUSION

In order to detect the regions of activity in the six hand gestures considered in this work, a probabilistic analysis of their sEMG signals is performed by using HMM. The detected states for each instance generate the predicted sequence which is compared to the initial stimulus at the transition regions for each subject. In order to validate the results at these regions, SVM classification of the detected outputs is carried out and compared to the classification of original RMS values. The classification accuracies improve by around 10% for the proposed HMM-based activity detection algorithm as compared to the original stimulus. The proposed process suggests ample scope in biomedical gesture analysis and development in diagnostic learning.

ACKNOWLEDGMENT

The authors would like to recognize the funding support provided by the Science & Engineering Research Board, a statutory body of the Department of Science & Technology (DST), Government of India, SERB file number ECR/2016/000637.